# Waveform cross correlation for seismic monitoring of underground nuclear explosions

## Part I: Grand master events

Dmitry Bobrov, Ivan Kitov, and Mikhail Rozhkov

Comprehensive Nuclear-Test-Ban Treaty Organization

**Abstract**

Seismic monitoring of the Comprehensive Nuclear-Test-Ban Treaty using waveform cross correlation requires a uniform coverage of the globe with master events well recorded at array stations of the International Monitoring System. The essence of cross correlation as a monitoring tool consists in a continuous comparison of digital waveforms at a given station with waveform templates from the global set of master events. At array stations, cross correlation demonstrates a higher resolution because the time delays at individual sensors from master and slave events are the same but they may differ from theoretical ones used in standard beamforming. In the regions where master events and thus waveform templates are available, one can reduce the amplitude threshold of signal detection by a factor of 2 to 3 relative to standard beamforming and STA/LTA detector used at the International Data Centre. The gain in sensitivity corresponds to a body wave magnitude reduction by 0.3 to 0.4 units and doubles the number of detected events. This gain is crucial for seismic monitoring under the CTBT. The coverage by real master events is sparse and confined to areas with historical seismicity, however. In two parts of this study, we investigate the possibility to populate the global grid with real and synthetic master events. In Part I, we replicate a high-quality master event over a regular grid several hundred kilometers from its actual position. In Part II, we model waveform templates using synthetic seismograms with the aim to apply them in aseismic zones. Both approaches are tested using the aftershock sequence of the April 11, 2012 Sumatera earthquake (Ms(IDC)=8.2). We used sixteen master events to recover the aftershocks in the Reviewed Event Bulletin of the IDC. The cross correlation standard event list built by these sixteen masters is a natural benchmark to evaluate the performance of the replicated masters and synthetic waveforms. The replicated real masters demonstrate the performance at the level of real masters.

# 1. Introduction

The International Data Centre (IDC) of the Comprehensive Nuclear-Test-Ban Treaty Organization (CTBTO) is obligated to find and report on seismic events, which, with given quality criteria, are possible to detect and build using the International Monitoring System (IMS). The IDC shall progressively develop the procedures for the production of standard reporting products and for the performance of standard range of services for States Parties (Protocol to the CTBT, Part 1, §17). Waveform cross correlation has lately demonstrated the possibility of a fundamental improvement in detection, location, and magnitude estimation when applied to large historical datasets. The IDC database, which includes more than 450,000 seismic events and tens of millions of raw detections, is a natural candidate for an extensive cross correlation study and the basis of further enhancements in monitoring capabilities. Without the historical dataset, which includes interactively reviewed events, this study and any improvements would not be feasible.

The IDC has been collecting and archiving a high quality set of uninterrupted seismic data from the primary IMS stations since 2000. When complete, the IMS includes fifty primary seismic stations. By design, they are distributed over continents (except island station PPT) with a general goal to have at least three primary IMS stations at regional distances from any point within continents (see Figure 1). Figure 1 also illustrates the well-known observation that global seismicity is highly inhomogeneous in space. A few relatively small areas contribute more than 95% of all seismic events built by the IDC. The geographic distribution of seismic events in the Reviewed Event Bulletin (REB), which is the final product of interactive human analysis at the IDC, is also highly inhomogeneous.

Waveform cross correlation improves detection (e.g. Gibbons and Ringdal, 2006, 2012; Gibbons *et al.*, 2012; Harris, 2006, 2008; Harris and Pike, 2006) and enhances phase association and event building (e.g. Harris and Dodge, 2011; Bobrov *et al.*, 2012abc; Slinkard *et al.*, 2012).

The use of cross correlation for monitoring purposes has a severe limitation, however. It is confined to the areas with historical seismicity, where master events and high quality waveform templates are available. This confines cross correlation to regional studies. For example, Geller and Mueller (1980), Schaff and Waldhauser (2005, 2010), and Waldhauser and Schaff (2008) applied cross correlation to seismicity in California. Schaff and Richards (2004, 2011) studied repeating events in China. Israelsson (1990) used cross correlation to characterize close regional events in Sweden, Joswig and Schulte-Theis (1993) analysed the mining-induced seismicity in the Ruhr basin of NW Germany, and Gibbons *et al*. (2007) detected and located earthquakes in northern Norway. At the same time, teleseismic and global studies are not so frequent and are chiefly associated with relocation/tomography exercises (e.g. Vandecar and Crosson 1990; Waldhauser and Schaff, 2007; Pesicek *et al*., 2010).

The limitation in spatial distribution of master events is especially discouraging for the cross correlation technique in view of excellent detection and location capabilities it has demonstrated when applied to three underground tests conducted by the DPRK, which were announced as nuclear explosions. Using the cross correlation detections at a few high quality IMS arrays at teleseismic distances, Selby (2010) located the 2009 DPRK test in an opposite direction relative to the REB location: ~2 km to the west and slightly to the north of the 2006 test. This location is supported by satellite images and accurate relocation with regional stations (Wen and Long, 2010). Therefore, cross correlation can provide a more accurate location compared to the standard locator used by the IDC. The overall improvement in the detection threshold associated with the DPRK events was exercised by Gibbons and Ringdal (2012). They presented a cross correlation detector for IMS station MJAR based on the 2006 DPRK test as a template, which allows lowering the detection threshold within the DPRK test site to magnitude 3.0, retaining a negligible false alarm rate. Bobrov *et al.* (2012b) also estimated this detection threshold between 2.5 and 3.0 using cross correlation at regional and teleseismic IMS stations.

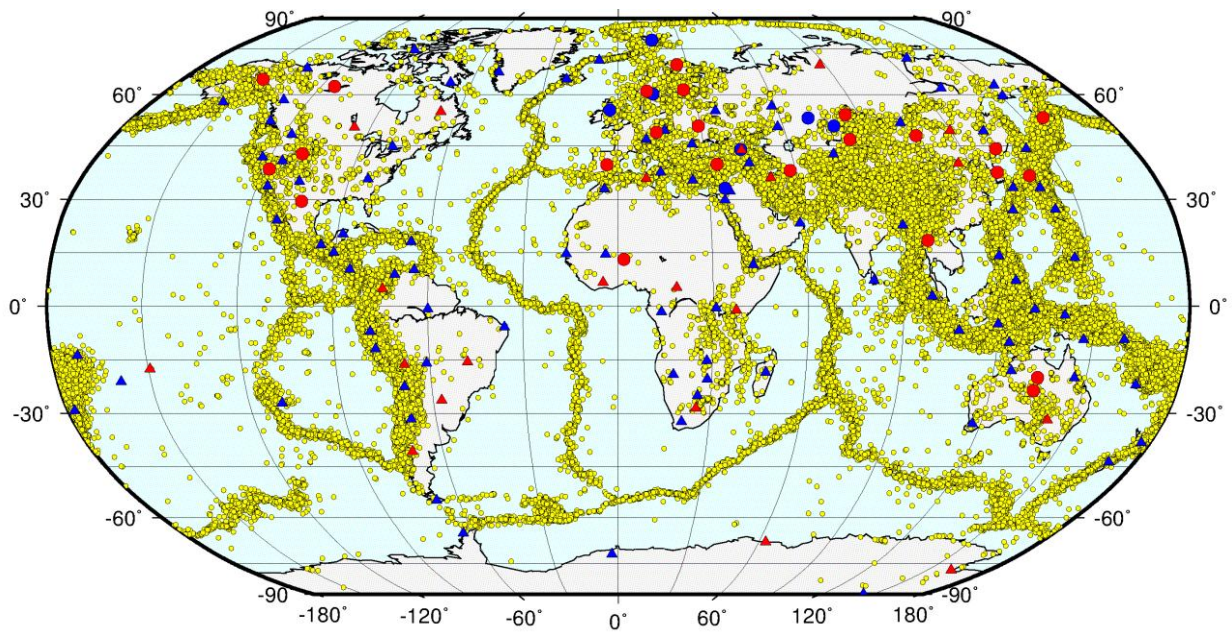

Figure 1. The IMS network includes 50 primary seismic stations, which are divided into seismic arrays (red circles) and three-component (3-C) seismic stations (red triangles). (The location of one primary station is not defined yet.) Auxiliary seismic arrays (blue circles) and 3-C stations (blue triangles) are also shown. Only seismic stations operating during the studied period are displayed. are. The REB includes >450,000 events with seismic phases. Those 350,000 having depths less than 50 km are shown by yellow circles.

For global monitoring, the backbone of cross correlation is a set of master events (earthquakes or explosions) with high quality template waveforms at IMS array stations. The array stations are most efficient for cross correlation because of destructive interference of all signals having vector slowness different from that of a given master event. These master events have to be evenly distributed and their template waveforms should be representative (similar in shape to most of waveforms from the same area) and pure (negligible noise input). The coverage and characteristics of historical seismicity observed by the IMS seismic network since 2000 does not match these requirements. The REB allows selecting a number of master events in seismically active areas but even there the quality of templates varies from master to master. Instead of using local master events we propose to replicate the best master events, which are called "grand masters", over a regular grid expanding several hundred kilometers from their epicenters. For a grand master event, the templates at IMS array stations have empirical time delays between individual sensors corresponding to the azimuth and slowness for the grand master/station pair. For its replicas, all time delays are corrected for the relevant theoretical difference between the original and displaced positions. The empirical deviations from the theoretical arrival times at individual sensors are inherently related to seismic velocity structure beneath the station. These empirical residuals are retained for all replicas even when shifted by several hundred kilometers. Despite being small, these residuals play the key role in the effectiveness of cross correlation as applied to weak signals. They define the advantage of cross correlation over beam forming, where all channels are stacked using theoretical delays. This advantage has been proved by a large number of studies at local, regional, and teleseismic ranges (e.g. Gibbons and Ringdal, 2006; Waldhauser and Schaff, 2007). To this end, we use only IMS arrays in our analysis.

Using waveform cross correlation, we have studied several aftershock sequences: after a small earthquake in the North Atlantic (Bobrov *et al*., 2012a), a middle size event in China (Bobrov *et al*., 2012b), and that after a magnitude $M_s$(IDC)=8.2 earthquake near Sumatera (Bobrov *et al*., 2012c). The number of new (in addition to the official REB) aftershocks in the cross cross correlation standard event list (XSEL) matching the IDC event definition criteria (EDC) varies from tens to several hundred, but comprises almost the same portion, 50% to 70%, of the number of aftershocks reported in the REB for the same events. To find new events we used various sets of master events – from just one master event from the REB to all events in the REB together with all newly built events.

For the earthquake in the North Atlantic, we used all aftershocks as master events. In total, the cross correlation detection and phase association techniques found all 38 REB events and 26 new events. The latter had all formal properties of the REB events. There were several REB events with $m_b$(IDC) above 4.0. These aftershocks were the most efficient master events in terms of the number of found REB and new earthquakes. The best master with $m_b$(IDC)=4.25 built 54 events from the total of 64. These masters have a relatively large number of array stations and their waveform templates are characterized by SNR above 5.0. This is a mandatory template quality threshold.

The March 20, 2008 Chinese earthquake sequence was recovered with one, ten, thirty and 120 masters. Due to limited human resources, there were only 45 from 120 hypotheses reviewed by experienced IDC analysts with 36 valid REB ready events built. All bigger aftershocks in the Chinese sequence can be used as masters - they were close in space and thus could effectively replace each other. This was a crucial finding for the global grid with masters separated by 50 to 70 km.

At regional distances, almost all aftershocks were measured by IMS station MKAR. The Pn-wave cross correlation detector showed the same efficiency as that based on teleseismic P-waves, with the correlation distance of tens of kilometers. We used a 10-second correlation window for the Pn-wave. All regional studies based on longer templates for the Pn and Lg-waves, which provide a high level of specificity of the sought waveforms, report the master/slave correlation distances of a few kilometers (e.g. Nakahara, 2004; Baisch *et al.,* 2008).

For the Sumatera earthquake, we had to introduce sixteen master events from the earlier aftershocks in order to cover the whole aftershock area of 500x500 km. This case is considered as a small-scale prototype of the global master coverage and highlights major problems with master event selection and conflict resolution between event hypotheses created by adjacent masters. Because of the number of master events and the area of the aftershock sequence this case is used as a benchmark in order to assess the performance of grand masters. Basically, we compare the number of detections, associated arrivals, and built events in various configurations of master events. The overall study is split into two major parts. In Part I, we replicate the best event from sixteen real masters. There are two replication schemes: over the actual position of real master events and over a regular grid. In Part II, we model waveform templates using synthetic seismograms with various source and velocity models. If cross correlation based on synthetics performs well without prior information on source functions and velocity distributions one can apply them in aseismic zones.

2. **Data, cross correlation, event building, conflict resolution, and grand masters**

For comprehensive seismic monitoring of underground nuclear explosion (UNE), the smallest events are most important. They are more likely to be missed by standard detection algorithms and event building tools used at the IDC. Therefore, the template windows should not be lengthy and should only include valid signals from small and moderate-size events. Numerous historical and most recent observations show that the signals generated by underground nuclear explosions are generally of impulsive character and fade away after few oscillations. Longer and complex P-waves from UNEs are less frequent and do not compromise the choice of short templates.

Our experience has shown that the most effective length of correlation window is 4.5 s to 6.5 s depending on the frequency band (Bobrov *et al*., 2012abc). For a standard regional array with 10 individual instruments and aperture of 4 km, these time windows balance the duration of weakest signals and the azimuth/slowness resolution. It is worth noting that the total length of the cross correlation templates is 45 s to 65 s, and thus, they are very specific for master/station paths. Shorter windows would likely fail to accurately estimate the signal vector slowness using standard *f-k* analysis. The templates lose their specificity and the array would detect signals from various directions. Broader correlation windows involve too much microseismic noise suppressing the CC estimates for the shortest signals, which are expected from low yield tests or those conducted in large underground cavities. Taking into account the range of variation in frequency content of the ambient microseismic noise and the signals of interest for seismic monitoring, which depends on the source/station configuration and varies with time, one requires a number of filters covering the spectral range from 0.8 Hz to 6 Hz. We use four frequency bands standard for IDC detection: 0.8 Hz to 2.5 Hz, 1–3 Hz, 2-4 Hz, and 3 to 6 Hz (Coyne *et al.*, 2012). Shorter windows correspond to higher frequencies in order to retain the total number of oscillations in the templates.

To calculate cross correlation coefficient, CC, we average individual normalized cross correlation functions over all channels, as proposed by Gibbons and Ringdal (2006). This allows avoiding many problems with data quality such as spikes, gaps, high noise at individual channels. When one or a few channels have unrecoverable quality problems we apply a taper with the channel weight falling from 1 to 0 in the problematic interval to suppress the input of this (-ese) channel (-s). As a result, the averaged cross correlation trace has no artificial steps associated with a sudden change in the number of channels which might be interpreted as a valid detection. For the global network of 50 stations with a multitude of natural, technical, and human factors affecting the performance of numerous individual channels this technique is a more reliable one than two alternative methods of CC calculation. Firstly, instead of cross correlating individual channels, when all of them are shifted relative to the reference station by theoretical time delays defined by the master/station vector slowness, it is possible to concatenate all individual waveform segments in a given template window with the relevant time shifts and to calculate CC. This method is subject to quality problems described above. Secondly, one can cross correlate beams of the master template and continuous waveform, both steered to the master. This method is inferior because the estimate of CC is subject to the same data quality problems and also to the beam loss associated with the deviation of actual time delays from their theoretical predictions due to non-planar P-wave propagation across the array in highly inhomogeneous velocity structures beneath some arrays.

When an aggregated time series of cross correlation coefficients is calculated for a given interval, signal detection algorithms are applied. It is not possible to apply a simple amplitude detector to original waveforms because of strong variation in the level of microseismic noise. Since |CC| varies between 0 and 1, the absolute detection threshold has a clear sense for the CC trace. However, there is no global and unique |CC| threshold, $CC_{tr}$, to define a new arrival. These thresholds are station dependent and vary with geographical coordinates. In this study, we required all valid detections to have |CC|>0.2.

The similarity in source functions depends on local velocity structure, depth, source mechanism and also may deteriorate with the difference in magnitude, especially for short time windows used in our templates. Shallow earthquakes may generate emergent signals. For larger events, an early (and different in shape) part of the signal from the same location may emerge from the noise, and thus, be used in corresponding templates. For smaller events, this initial part is often well below the noise level. As a result, the level of cross correlation may drop below the threshold even for collocated events with a large difference in magnitude.

For weak signals, the absolute level of correlation coefficient for collocated events can be reduced by the effect of uncorrelated seismic noise mixed with the signals. Hence, before using CC as a detector, one has to enhance the detection procedure. There are many possibilities and likely the simplest one is the STA/LTA detector, which is already implemented at the IDC for original waveforms. This detector is based on a running short-term-average (STA) and long-term-average (LTA), which is computed recursively using previous STA values (Coyne *et al*., 2012). The LTA lags behind the STA by a half of the STA window. The length of the STA and LTA windows have to be defined empirically as associated with spectral properties of seismic noise and expected signal. We have been experimenting with both lengths and determined two optimal windows: 0.8 s for the STA and 20 s for the LTA (Bobrov *et al*., 2012b).

A valid signal is detected when the level of STA/LTA (SNR_CC) is above 2.5. This is a tentative but a conservatively low threshold. Before we gather a statistically significant set of arrivals it would be premature to increase the detection threshold. It is worth noting that for original waveforms a valid signal usually has SNR>2.0, but our CC detector can find a valid signal with standard SNR=0.7 (Bobrov *et al*., 2012b).

Generally, a higher cross correlation between two waveforms at one station is a reliable indication of the spatial closeness between their sources. However, there are a few cases when cross correlation is high for distant events, with the master event being much smaller than that found by cross correlation. There are several methods to remove such spurious correlations. One can consider these methods as additional filters applied to the flux of detections used for event building. IDC automatic processing uses *f-k* analysis applied to the original waveforms in order to estimate the difference in slowness and azimuth between two sources. When this difference is high one can reject the null hypothesis that these sources are close. The overall resolution of *f-k* analysis is not very high, especially, for small events generating weak signals.

Gibbons and Ringdal (2006) proposed and successfully applied *f-k* analysis to the cross correlation time series at array stations. This allowed a significant improvement in the overall resolution due to the sensitivity of correlation to the distance between events and effective noise suppression. We also estimated pseudo-azimuth and pseudo-slowness using *f-k* for correlation time series for all relevant detections. (We use term "pseudo" since there is no one-to-one correspondence between the true and CC domains, and the estimated azimuths and slownesses are not expressed in genuine degrees and seconds per degree.) This procedure allows effectively rejecting most of cross correlation detections associated with noise and remote events (Gibbons and Ringdal, 2012; Bobrov *et al*., 2012b).

There is a dynamic method to sort out all inappropriate arrivals (Bobrov *et al*., 2012b). It is based on the ratio of signal norms: $|\mathbf{x}|/|\mathbf{y}|$, where $\mathbf{x}$ and $\mathbf{y}$ are the vector data of the slave and master, respectively. The logarithm of the ratio, $RM = \log(|\mathbf{x}|/|\mathbf{y}|) = \log|\mathbf{x}| - \log|\mathbf{y}|$, is the magnitude difference between two events or relative magnitude. This difference has a clear physical meaning for close events with similar waveforms. For a given slave event, the relative magnitude is a reliable dynamic parameter for a correct arrival association at several stations. For two events with close locations, the level of cross correlation coefficient at a given array station with a fixed configuration of individual sensors depends on the distance between events, the similarity of source functions, the difference in velocity/attenuation structure along propagation paths, and the change in spectral characteristics of microseismic noise. Apparently, for larger master/slave distances CC is lower. Together with decreasing CC, the offset results in the travel time change measured at the reference channel as well as in the relative travel times to other sensors of the array. For digital waveforms, the change in relative arrival times between the channels is not continuous. It has a characteristic offset, which depends on the rate of digitization, the distance between sensors, and the difference in vector slowness in the slave/master pair. For a given channel, the arrival time shift, *dt*, relative to the reference channel (which is usually located near the geometrical centre of the array) can be calculated according to a simple relationship:

$$dt = \mathbf{s} \cdot \mathbf{d} \quad (1)$$

where $\mathbf{s}$ is the vector slowness of the P-wave arrival at the reference array station, and $\mathbf{d}$ is the vector between the reference and given sensors. For a digitization time step, $\Delta T$, there is no time shift relative to the reference channel when $dt < \Delta T$, i.e. both digital waveforms have the same arrival time. If the largest $dt$ at a given array for a source at a teleseismic distance is less than $\Delta T$, then all arrivals on individual channels are effectively synchronized.

Using (1), we can determine the distance between a master and a slave event, which results in a resolvable change in relative arrival time. In an array, let's define the largest distance among all sensors to the reference channel in the direction of the master event as $\mathbf{d_1}$ and that in the direction of the slave as $\mathbf{d_2}$. The vector slownesses at the reference channel from the master and slave are $\mathbf{s_1}$ and $\mathbf{s_2}$, respectively. Then the difference in arrival times at these remote channels is

$$dt_1 - dt_2 = \mathbf{s_1} \cdot \mathbf{d_1} - \mathbf{s_2} \cdot \mathbf{d_2} \tag{2}$$

Taking into account that for close master and slave events the most remote channel is likely the same ($\mathbf{d_1}=\mathbf{d_2}$) one obtains from (2) a simple inequality for the resolvable change in arrival time, $\delta t = dt_1 - dt_2$:

$$\delta t = (\mathbf{s_1} - \mathbf{s_2}) \cdot \mathbf{d_1} > \Delta T \tag{3}$$

From (3), one can always estimate the characteristic master/slave offset for a given master/station pair. Within this characteristic radius from the master event, any slave position is characterized by the same time shifts between individual channels. This radius defines the smallest spacing between masters for a predefined station set. When a grand master is replicated over a regular grid, no two masters should be closer than the smallest characteristic distance amongst all involved stations. Otherwise, there is no change in the cross correlation coefficients between adjacent replicated masters with identical waveform templates.

For each master event, a set of IMS array stations should be defined with template waveforms matching several quality criteria. For the purposes of nuclear tests monitoring under the CTBT, the task of the IDC is to find, build and estimate parameters for all events, which match the IDC event definition criteria (Coyne *et al*., 2012). Using the template waveforms, we calculate cross correlation coefficients for individual channels, average them to obtain an aggregate CC-trace for each station and then apply detection procedures as based on |CC| (>0.2) and SNR_CC (>2.5). For all detections, we estimate pseudo-azimuth and pseudo-slowness using the calculated CC-traces. After all arrivals with azimuth and slowness residuals beyond ±20° and ±2 s/deg, respectively, are removed, one has a set of detections with their arrival times for each station, $t_{ij}$, where $i$ is the index of the $i$-th arrival at station $j$. For the events close to the master, the travel times to all relevant stations can be accurately approximated by the theoretical master/station travel times, $tt_j$. We use the IASPEI91 velocity model. Using the model predicted travel times to the master and the measured arrival times one can calculate the approximate origin times, $ot_{ji}$, for all detections:

$$ot_{ji} = t_{ij} - tt_j \qquad (5)$$

Therefore, the set of arrival times is converted into a set of origin times for an unknown number of hypothetical slave events. By definition, an origin time is a genuine characteristic of source. According to the EDC, we need three or more stations to detect dynamically consistent signals with origin times within a few seconds, which is an equivalent to travel time residuals. Then, it is possible to associate these arrivals with a unique event. For this event hypothesis, we average all associated origin times and assign the estimated value to the event origin time.

Different from IDC automatic processing, there is no need to locate the built event since it has to be within the master's origin neighbourhood. It should be noticed that the double difference location based on the differential travel times estimated using cross correlation generally needs more than three stations for a reliable solution. The distance between two events with a high cross correlation coefficient between their waveforms may reach 50 and more kilometres, as relationship (5) suggests. For a slave at 50 km, the master travel time is not a good approximation. In order to improve the process of origin time association with a unique slave event we have introduced an equidistant mesh around the master event, which was implemented in a form of two circles of 25 and 50 km in radius. There were 6 and 12 mesh points distributed over two circles, respectively. In total, each master gives birth to 19 nodes. For each node, all arrival times are reduced to origin times using the theoretical travel times corresponding to the node location. When the arrival times are accurately estimated, the search over 19 nodes provides the smallest RMS origin time error for the mesh point likely closest to the sought slave event. It is worth noting that, according to (5), for all 19 nodes the cross correlation coefficient between the master and slave waveforms is the same. Actually, the cross correlation distance may be more than 100 km (Bobrov *et al*. 2012c), but we limited it to 70 km because the distance between the real masters for the Sumatera aftershock sequence is less than 100km. We could also use a denser mesh for a better location, but this issue is beyond the scope of the current study. In any case, the 25 km spacing allows an approximately 12.5 km location accuracy, which corresponds to the confidence ellipse of ~500 km$^2$.

We call the process of origin time association the "local association", *LA*, in line with the name of global association, *GA*, used by the IDC. Indeed, only the phases from events local to the master ones are associated. The *LA* does not see any events beyond the radius of correlation. When a distant event (say, beyond 100 km) demonstrates good cross correlation coefficients with a given master event, the origin times obtained from the arrival times at the master-related stations scatter beyond the predefined limits. In this case, no event hypothesis can be built.

Here we assume accurate arrival time estimates and good station coverage with a reasonably small azimuthal gap. The azimuthal gap issue can be resolved by the introduction of a specific threshold for a given master since all stations with waveform templates are known in advance. For example, in the study of the 2008 aftershocks in China we excluded the hypotheses created by any three-station combination of FINES, ARCES, GERES, and NOA. The accuracy of arrival time estimates obtained by cross correlation is less accurate for very weak signals. The influence of noise may shift the onset time only by few seconds.

We have started with a six-second time window in the *LA* for the association into a single event. Considering the travel time uncertainties in the GA (tens of seconds), this is a very short interval for origin times. It corresponds to the difference in travel times between the master event and an event on the rim of cross correlation zone. Disregarding the mesh around each master event, which takes into account the change in travel times, it is possible to estimate the scatter of origin times for a master/slave distance of 50 km. For a teleseismic P-wave with 0.05 s/km slowness, the travel time difference is of 2.5 s. For two stations in opposite directions, two 2.5 s travel time residuals give a 5 s difference in origin time. This is the worst case scenario. A 1 s uncertainty in onset time may add 2 s to the origin time difference. We have tested longer windows and found six seconds to provide almost as mjany events as a nine second window. For the latter window, more noise phases might be wrongly associated and additional events are likely less reliable. The length of the origin time window is of crucial importance only for the aftershock sequences of catastrophic earthquakes with a hundred events per hour. For the rate of a few events per year, there are no competing hypotheses created by the *LA*.

There is an important enhancement of the *LA* process based on the relative magnitudes. All arrivals associated with a given event should have *RM* estimates within some predefined bounds separating the true and bogus signals. We have adopted a tentative value of 0.7, which is much smaller than a similar magnitude difference of 2.0 used in IDC automatic processing. The 0.7 magnitude difference corresponds to the deviation of master/slave amplitude by 5 times from the network average, while the difference of 2.0 corresponds to a factor of 100. The 0.7 threshold is then tested on the full set of found events.

The *LA* is a simplistic process compared to the global association. A big advantage of the *LA* is a significantly reduced number of arrivals for a given master event from the associated stations. At the same time, the total number of arrivals at a given station may double relatively to that from the current IDC detector. We have already mentioned that the cross correlation detector can find valid signals with SNR<1.0. This increased number of arrivals, however, can be effectively split into a large number of subsets associated with different masters. All masters are processed independently.

When several master events are close in space, their templates may correlate with waveforms from the same event, and thus, create similar new event hypotheses with very close arrival times at several stations associated with these masters. To select the best one from a few similar events with close arrival and origin times we count the number of stations used in these hypotheses. If one event has the largest number of stations it is retained as an XSEL event. When several events have the largest number of stations we select the one with the smallest origin time standard deviation. By definition, this event is the most reliable and its parameters are written into the database. All competing hypotheses are rejected. Thus, for a multiple set of master events, the *LA* provides a unique set of found events. In our previous studies (Bobrov *et al*., 2012ab), we used the cumulative CC, $\sum|CC_j|$, to select the best event. This approach gave too much weight to misassociated phases. These wrongly associated arrivals actually increase the origin time scattering, but not the reliability of the hypothesis.

Figure 2 demonstrates various sets of master events for the Sumatera aftershocks. In the original study (Bobrov *et al*., 2012c), sixteen real masters were not distributed homogeneously in space and magnitude. Moreover, the SNR averaged over seven array stations used for cross correlation (SNR_AV) varied from 14 to 208. The difference in template quality and inhomogeneous distribution of aftershocks resulted in a varying number of event hypotheses built by each of sixteen masters: from 1748 to 2636 for the period from April 11 to May 25. We have found a logarithmic dependence between the number of event hypotheses and the SNR_AV, as shown in Figure 3.

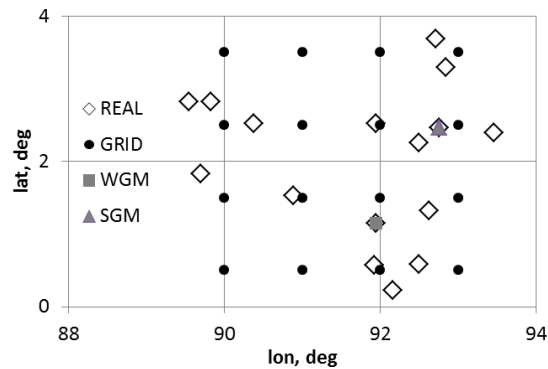

Figure 2. Location of sixteen real master events (open diamonds), the SGM (grey triangle), the WGM (grey square), and the regular grid with 1$^o$ spacing (black circles).

The detection and event statistics previously obtained for the Sumatera aftershock sequence allows a direct comparison of the real and replicated master sets. In this study, we replicate sixteen real master events with one best grand master. We call this event "strong grand master" (SGM). This replacement may increase the correlation coefficient for those master/slave pairs, where the real master template includes more microseismic noise than the grand master. As a benefit, the grand master can be selected to have the largest possible number of qualified (SNR>5.0) stations, while the real masters have from seven to sixteen stations. Since we replicate the grand master over the whole aftershock zone, it is straightforward to compare the result of the real masters with seven stations and those obtained using the grand master with sixteen station. One can expect more arrivals to be detected, and thus, a larger number of event hypotheses with the sixteen station masters. An opposite effect may arise from the difference in focal mechanism/source function between real and grand masters. In some places, the grand master might be not a representative one and suppress their performance relative to real master. Overall, if the replicated masters build more valid hypotheses, the cross correlation technique actually gains from the replacement.

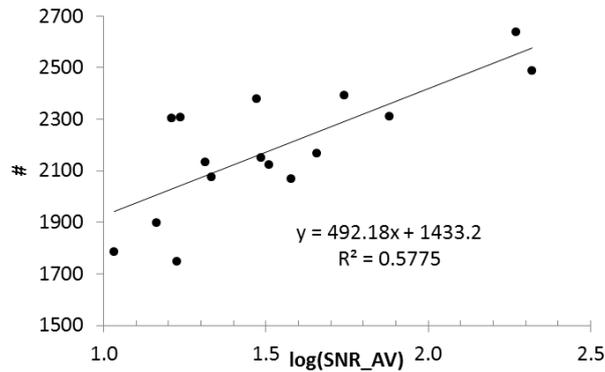

Figure 3. The number of built events as a function of the logarithm of the SNR_AV for sixteen real masters.

Another possibility to improve the overall performance of cross correlation is to create a regular grid instead of actual distribution of historical events, as shown in Figure 2. This approach has two important advantages: the location accuracy and computer resource reduction. The former critically depends on the number of defining stations and the precision of arrival times. Apparently, the best master is characterized by the highest location accuracy. When replicated over a regular grid, this master has the same location accuracy for each node. This also guarantees controlled coverage and detection sensitivity for the whole studied area without lacunas associated with actual earthquakes taken as masters. Figure 4 illustrates the distribution of additional grid points used for travel time calculations around the nodes of the regular grid.

A regular grid also reduces calculations. The grid masters retain the empirical time delays at distances of several hundred kilometers. For a master event replicated over an area 1000x1000 km$^2$ with 100 km spacing the cross correlation calculations are reduced by two orders of magnitude. One can also introduce a denser grid not using extra computer resources.

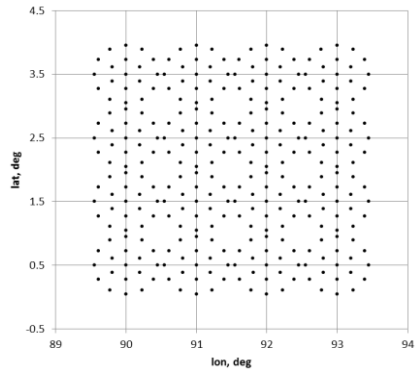

Figure 4. The mesh used in local association. Each of sixteen grid points is surrounded by two rings of 6 and 12 nodes at distances 0.225$^o$ and 0.450$^o$, respectively. For a given master, the *LA* creates 19 event hypotheses by reducing all arrival times to origin times using the relevant travel times.

Having two configurations for the grand master distribution, it is instructive to evaluate the influence of its quality, which is characterized by the average SNR. We have chosen the poorest master or "weak grand master" (WGM) and tested its performance. In all cases, the studied period was limited to 24 hours after the main shock. This is the most difficult period for detection and event building. At the same time, it guarantees an extensive statistics and reliable inferences.

3. **Comparison of real and grand masters**

The applicability of the grand master approach has to be proven statistically. We have to compare several sets of arrivals and XSEL events obtained by sixteen real master events and by a grand master replicated over the locations of real masters and over a regular grid. This small grid of replicated grand masters is considered as a part of the global grid. Therefore, its performance is crucial for the design of IDC automatic processing based on cross correlation.

All stages of cross correlation processing should be assessed: detection, phase association and event building for each individual master, and conflict resolution. Comparison of the automatic XSELs to the Reviewed Event Bulletin is the final step. It should be noticed, however, that the REB misses many valid events and contains a few per cent of bogus events. These bogus events may be built from phases arrived from different sources or later phases of the same source. In the latter case, the events are partly bogus but mislocated by hundreds kilometers and their origin times are tens of seconds off their actual times. Hence, one should not rely on the REB as a perfect catalog to be repeated by the XSEL.

Bobrov *et al*. (2012c) introduced two reference cases: sixteen real masters with seven best station each and the same master events with varying number of stations, all having SNR>5.0. The latter case created clear difficulties for conflict resolution between adjacent masters and was not fully analyzed. At the same time, it builds the highest number of events and may serve as the high end benchmark. Here, we compare the results obtained with the best and the worst grand masters to both cases.

Table 1 lists total number of arrivals created in five master configurations: sixteen real masters (REAL), the SGM and WGM replicated over real master locations (SGM REAL and WGM REAL), and over the regular greed (SGM GRID and WGM GRID) shown in Figure 2. For the seven station case, the SGM in real locations detects more signals, events and arrivals associated with all built events than the real masters. The same situation is retained when all stations included in cross correlation processing, where the number of stations varies between sixteen real masters and is fixed to 17 for the SGM. This result does not contradict our intuition that better signals for sources in the same positions improve the estimates of cross correlation coefficient. In this regard, the signal quality factor may outperform the effect of variation in signal shape. For the global coverage by grand masters, a more important result is that the SGM on the regular grid has also created more detections and events that the set of real masters. This is not a direct proof that the grid is more effective than real locations in creating of REB events, however.

As expected, the WGM shows the worst performance with seven and all stations in both configurations. The gain associated with template quality may be expressed by the ratio of arrival numbers in seven-station cases: 18% and 14% for real and grid positions, respectively. The same ratios are obtained when all stations are used. The number of associated arrivals for the SGM is by a factor of 1.28 and 1.21 larger than that obtained with the WGM for seven and all stations, respectively. This also illustrates the quality of arrivals obtained by the SGM.

The number of events build by all masters is the highest for the SGMREAL case. This does not guarantee the best XSEL, however. The SGM produces more similar events which are rejected during the conflict resolution procedure. The REAL XSEL contains by 40% more events and by 34% more arrivals than the SGM REAL XSEL. Therefore, the REAL XSEL has fewer stations per event on average. The larger number of REAL XSEL events does not guarantee more new REB-EDC events.

The SGM demonstrates similar results for the real and grid positions. The number of XSEL events and arrivals differs by a fraction of per cent. This observation supports the effectiveness of the global grid creation using the grand masters approach. The overall similarity may vary between individual masters, however, and it is instructive to analyze their relative performance. When a real master performs better than the SGM, one can estimate the portion of missed events. These are the events we sacrifice in order to reduce the detection threshold on average. We have found that the portion of missed events is small in comparison to the overall gain associated with cross correlation. At this stage, we assumed that this portion is globally uniform and the XSEL will not miss more events in aseismic areas than missed in IDC automatic processing.

Table 1. Total number of arrivals, associated arrivals, and events build by different configurations of masters.

|  | Seven stations | | | | | All stations | | |
|---|---|---|---|---|---|---|---|---|
|  | arrivals | assoc. | events | arrivals, XSEL | events, XSEL | arrivals | assoc. | events |
| **REAL** | 166740 | 61354 | 15594 | 9631 | 2324 | 214478 | 89769 | 19262 |
| **SGM REAL** | 171535 | 64489 | 16256 | 7199 | 1669 | 218707 | 93529 | 19780 |
| **WGM REAL** | 144892 | 50547 | 13023 | 6050 | 1448 | 185823 | 74443 | 16302 |
| **SGM GRID** | 168361 | 62262 | 15779 | 7182 | 1659 | 215179 | 91077 | 19324 |
| **WGM GRID** | 147207 | 51374 | 13231 | 5839 | 1378 | 188274 | 75465 | 16566 |

Figure 5 depicts the relative change in the number of detections and events obtained by the SGM replacing sixteen masters in their positions. When all qualified stations for a given master are used for detection, the SGM_all/REAL_all curves show that the SGM performance varies around 1.0 for detections and built events. If to replace master #2 (WGM) with the SGM, the number of detections and events rises by a factor of 1.3. Obviously, master #2 is a poor choice in the original study. For master #12, the original set of templates creates by 20% more detections and events than the SGM it its position. This effect may be related to the predominant mechanisms and source functions of the aftershocks near the actual event. In the 7-station cases, SGM_7/REAL_7, the overall result is similar but the real masters with fewer stations improve their relative performance.

When all 17 stations in the SGM are used, the number of detections and events is always larger than for the 7-station real masters. The SGM outperforms the real masters and creates no problems with the varying number of stations for the conflict resolution. Hence, Figure 5 supports the use of replicated grand masters instead of real masters. The quality of the SGM detections and events has to be proven, however.

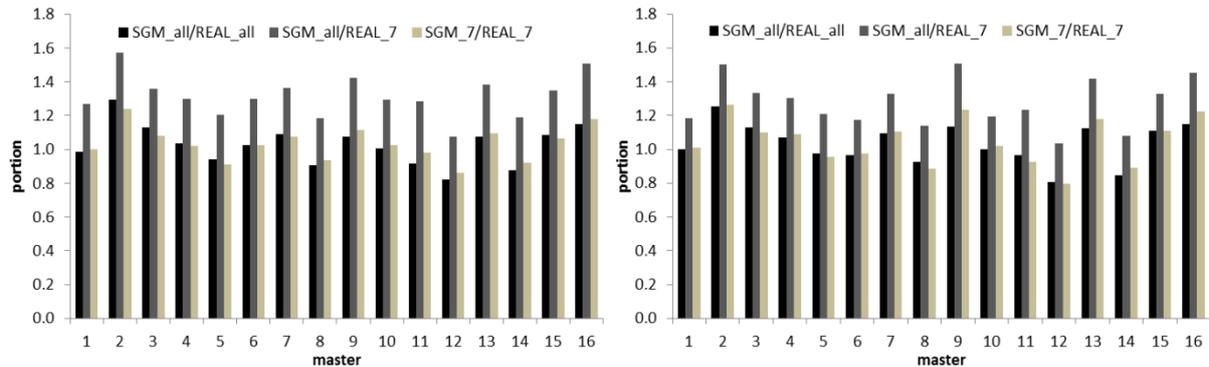

Figure 5. Left panel: The number of detections for the SGM replicated over the real master locations normalized to the number of detections obtained by the real masters. Three cases are considered: all detections obtained by 17-stations related to the SGM divided by the number of detections obtained by all stations, SGM_all/REAL_all; SGM_all divided by the number of detections obtained by 7-station real masters, SGM_all/REAL_7; and SGM_7/REAL_7. Right panel: Same as in the left panel for the number of events.

The effect of conflict resolution is best illustrated by the difference between subsequent origin times of the events found using the real and grand masters. Figure 6 displays six frequency distributions obtained in time bins defined by a geometric progression with common ratio 2 and scale factor 0.1. Before the conflict resolution, three curves peak near 1 s. There are plenty of events found by many masters. These small differences between the origin times are associated with the closeness of arrival times estimated using different masters at the same stations. This is a basis of effective removal of similar events, having arrival times within a few (4 in our study) seconds at the same stations. After the conflict resolution, three curves with extension "XSEL" peak at 25.6 s. Above 200 s, all six curves practically coincide. The events with large differences in origin times do not compete for the same arrivals and thus do not destroy each other. The "REAL XSEL" curve is higher than the other two XSEL curves between 1 s and 60s. This observation may reflect higher scattering of arrival times obtained by poor real masters. Then many similar events are not rejected in the conflict resolution procedure.

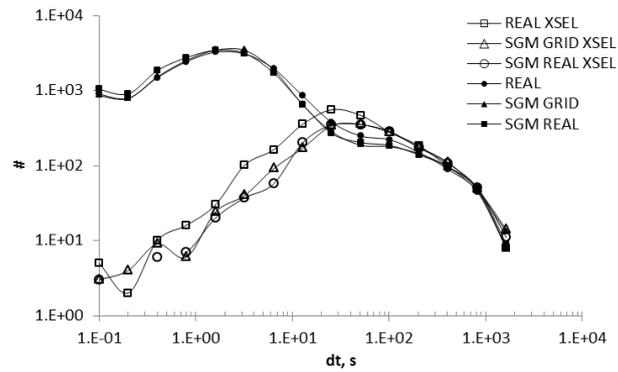

Figure 6. Six frequency distributions of the difference between subsequent origin times for 16 real masters and two SGM configurations (REAL and GRID) before and after conflict resolution. Notice the log-log scale.

Figure 6 also demonstrates that the conflict resolution procedure successfully removed 85% (REAL) to 90% (SGM) of similar events. Obviously, larger aftershocks with clear signals have a tendency to be found by many masters. This tendency may introduce a significant positive bias in the estimates of such event characteristics as the average and cumulative CC. Therefore, it is more reliable to use only the events and associated phases in the relevant XSELs for statistical inferences. Since we compare the real masters and the SGM, waveform templates at the following seven IMS stations are used: ASAR, CMAR, GERES, MKAR, SONM, WRA, and ZALV.

Figures 7 and 8 illustrate the quality of the XSEL events built by masters in five studied cases. The probability density functions of the event average *CC, CC_AVE,* demonstrate that the real masters built a larger portion of events with lower *CC*s. The SGM provides the highest *CC*s in both cases: for the real locations and when distributed over the grid. The cumulative *CC*, *CC_CUM*, i.e. the average *CC* times the number of stations, provides a complementary view. All curves practically coincide between 0.8 and 3.6 and peak at *CC_CUM*=1.4. In terms of the average and cumulative *CC*, the quality of events in all five cases is the same.

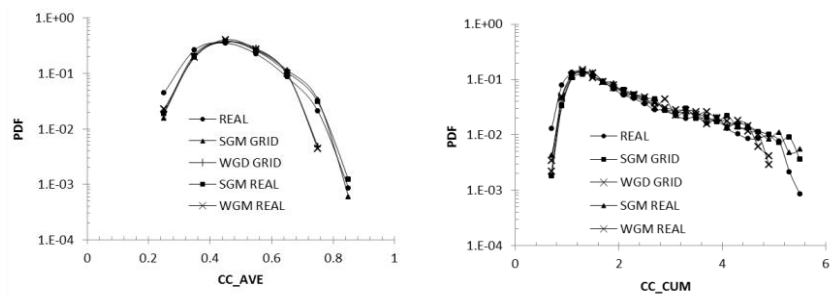

Figure 7. Probability density functions of the average *CC*, *CC_AVE*, (left panel) and the cumulative *CC*, *CC_CUM*, (right panel) for five cases. Notice the lin-log scales.

The distribution of relative magnitude in Figure 8 depends on the template RMS amplitudes. The SGM has the largest RMS amplitude at each of seven stations and the relevant *RM* values are lower than those obtained using the WGM. The difference between the SGM and WGM peaks is of 0.6, i.e. in a good agreement with their body wave magnitude difference of 0.56. All four grand master curves are characterized by a sharp fall at lower *RM*. The SGM has $m_b$(IDC)=5.06. Both related curves peak at *RM*=-1.4, and the lowermost *RM*=-2.4. Therefore, the SGM, either replacing the real masters or distributed over the grid, likely provides a complete event catalogue to $m_b$(IDC)=3.7. But it can hardly find any event with $m_b$(IDC)<2.7. The REAL curve contains masters of varying magnitude, which make its distribution smoother and broader. Some of these real masters are able to find aftershocks with $m_b$ by 0.2 units of magnitude lower than the SGM. Several events in the XSEL found by the SGM have *RM*~ 2.0, which makes magnitude 7.0. The REB does not contain events with so high magnitudes for the Sumatera aftershocks. The conventional body wave magnitude scale satiates at the level of 6.0 to 6.5.

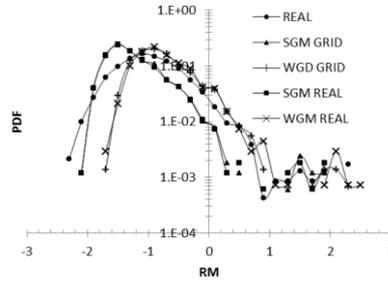

Figure 8. Probability density functions of the relative magnitude for five studied cases.

The quality of event hypotheses is defined by characteristics of associated arrivals. Waveform cross correlation provides two measures of signal quality: *CC* and *SNR_CC*. For a large aperture array with many individual sensors, even signals with *CC* of 0.1 can have *SNR_CC* above 3.0. Small arrays with a few sensors do not effectively suppress coherent noise components and even *CC* of 0.3 does not guarantee a clear signal. Figure 9 displays five PDFs estimated for *CC* and *SNR_CC* distributions in the relevant XSELs. The SGM demonstrates the same performance as the real masters. The REAL curve includes 16 autocorrelation cases which explain its deviation from the SGM curves near *CC*=1.0. As expected for the *CC_AVE* in Figure 7, both WGM curves do not contain large *CC*s.

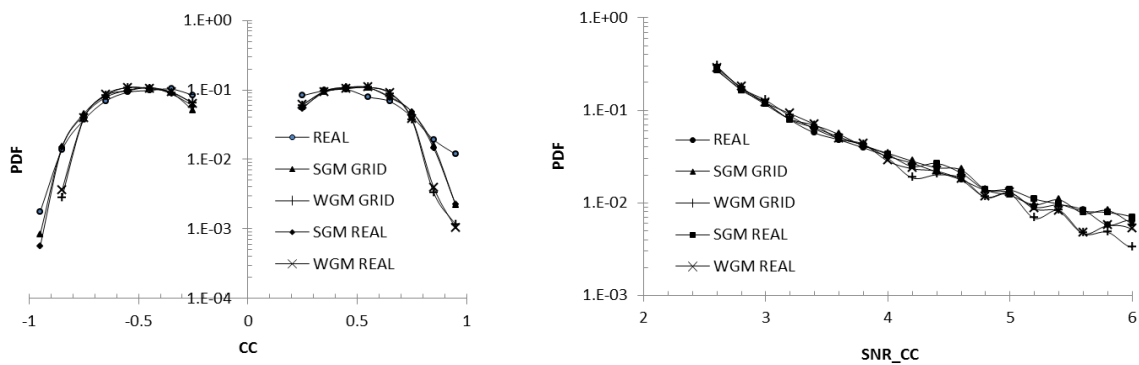

Figure 9. Left panel: Probability density functions of the *CC* at 7 stations for five cases. Right panel: Probability density functions for *SNR_CC*.

The distribution of station relative magnitude residuals, d*RM*, is displayed in Figure 10. The largest possible |d*RM*| is 0.7 and all PDFs are confined to this range. The SGM curves are characterized by lower levels at the wings, but both distributions are shifted by -0.1. This bias is likely associated with one or two stations, where the SGM templates have slightly higher RMS amplitudes than those predicted by its magnitude. Since all PDFs suffer a quasi-exponential fall from their peaks to ±0.7, the range was defined accurately and we do not miss appropriate arrivals in the XSEL events. This also suggests the absence of those bogus events, which could be wrongly built using dynamically inconsistent phases.

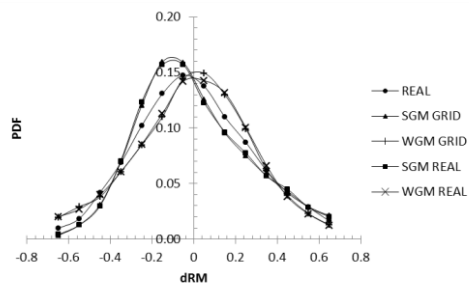

Figure 10. Five probability density functions of the d*RM*.

The cross correlation multichannel waveforms are used to estimate pseudo-azimuth and pseudo-slowness using standard *f-k* analysis. These estimates filter out all arrivals with azimuth and slowness deviating from their theoretical values by 20$^o$ and 2 s/deg, respectively. Figure 11 depicts five PDFs for azimuth, d*AZ*, and slowness, d*SLO*, residuals. The REAL d*AZ* curve has a sharper peak but all five d*SLO* curves are similar. Figure 11 indicates that, for a given phase, the probability to be associated with an XSEL event falls exponentially with the deviation in azimuth and slowness.

All curves in Figures 6 through 11 agregate the estimates obtained at seven stations. On average, the XSEL events and the associated phases detected using 16 real masters and the replicated SGM demonstrate quite similar features. There are some differences, however, which might be prominent at individual stations. Then, simple station corrections applied to the deviating parameters may substantialy improve the overall convergence. This is similar to the improvement in the uncertainty of network magnitude and origin time gained with station specific magnitude and travel time corrections. These are corrections for static errors.

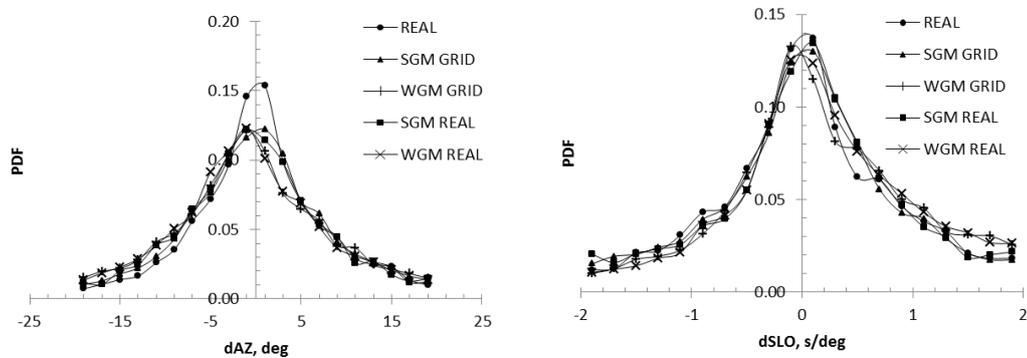

Figure 11. Left panel: Five probability density functions of d*AZ*. Right panel: Probability density functions of d*SLO*.

For seven IMS array stations, Figure 12 compares the d*RM* frequency distributions associated with 16 real masters and the SGM at the regular grid. Not surprizingly, these curves demonstrate larger differences and variations than those in Figure 10. For ASAR, the REAL curve is shifted by +0.1 to +0.2, has two peaks, and does not decay to the zero line at d*RM*=+0.7. The SGM curve is shifted by -0.1 and falls to zero at ±0.5. Hence, the SGM replicated over the grid is likely a better choice for ASAR than 16 different masters in terms of the relative magnitude estimate. For CMAR, the choice is opposite. The SGM curve is biased by +0.3 to +0.4 and the relevant XSEL likely misses many valid arrivals at CMAR with d*RM*>0.7. When a -0.35 correction is applied to all estimates of *RM* at CMAR, the SGM curve is shifted left with the peak at d*RM*=0. This correction, however, would change the network average defined by CMAR and other stations. There is a trade-off between the d*RM* corrections at seven stations, which is accompanied by varying numbers of associated arrivals. Therefore, these corrections should be calculated collectively and iteratively, with the aim of all individual curves to peak near d*RM*=0.

Stations GERES and MKAR demonstrate similar behavior, all curves are charaterized by peaks between -0.2 and -0.3. The SGM curve at station SONM is shifted by +0.3 to +0.4, with the REAL curve peak at 0. This difference also affects the relevant XSEL, but less than the bias in the CMAR estimates. For stations WRA and ZALV, the SGM curves are biased by -0.2, but both guarantee a low probability to miss an appropriate arrival due to d*RM*.

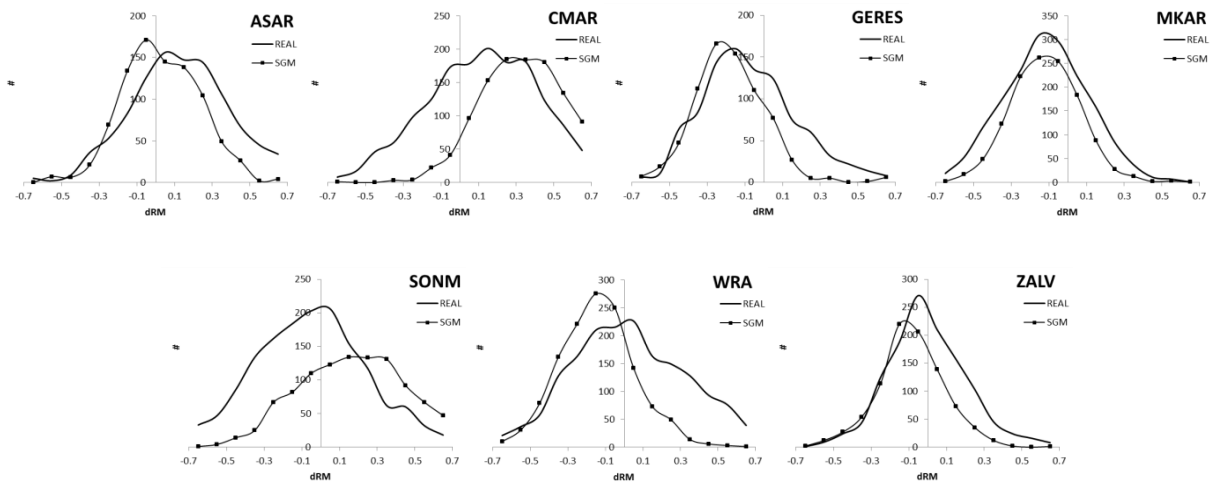

Figure 12. Frequency distributions of the station magnitude residuals as obtained from real master events (REAL) and the strong grand master replicated over a regular grid (SGM).

The frequency distributions of pseudo-azimuth and pseudo-slowness in Figure 13 also demonstrate strong variations in station performance. The d*AZ* and d*SLO* curves at large-aperture array WRA are all symmetric. They are decentered by a few degrees and fractions of s/deg, however. This effect is more prominent for MKAR (d*AZ*) and ASAR (d*SLO*). Since the azimuth and slowness estimates are obtained by *f-k* analysis the non-planar propagation of P-wave across a given array results in the difference between theoretical and measured values. This effect is well-known for arrays and is compensated at the IDC by slowness-azimuth-station-corrections (SASC). Not all IMS stations have these corrections yet. This is the reason why the REAL curve for MKAR is also biased. For the purposes of waveform cross correlation, we are going to calculate appropriate SASCs for all replicated grad master events. When applied, these corrections will return all d*AZ* and d*SLO* curves to the center. But they will not make these distributions sharper. The REAL and SGM curves at CMAR suggest that the estimates of azimuth and slowness for CMAR detections are definitely not the best for building XSEL events.

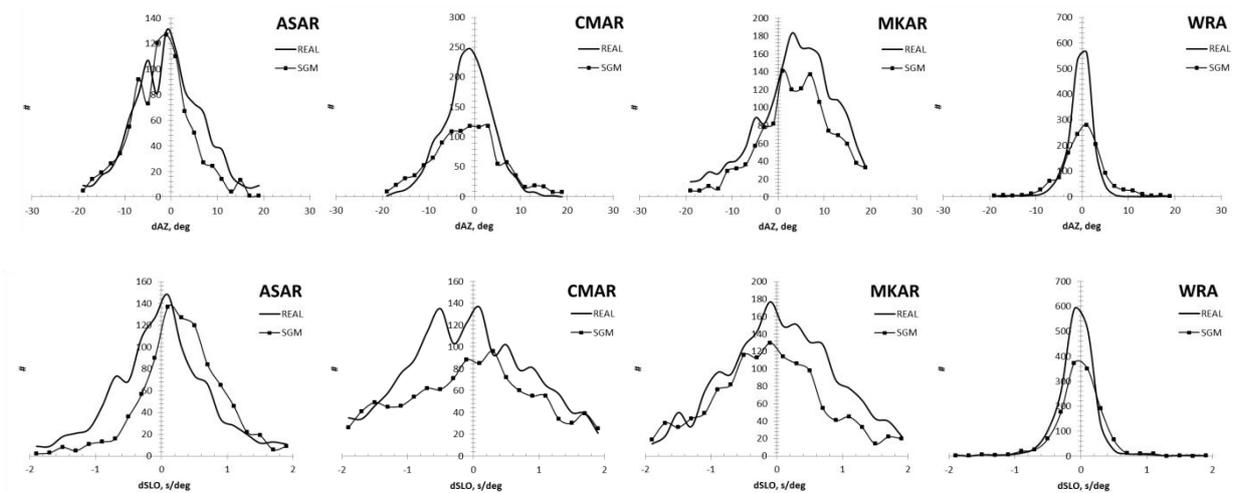

Figure 13. Frequency distributions of the residual azimuth, d*AZ*, and slowness, d*SLO*, for four IMS stations as obtained from real master events (REAL) and the strong grand master replicated over a regular grid (SGM).

The final comparison of three XSELs associated with 16 real masters and the SGM replicated over the real locations and regular grid is presented in Table 2. We checked the number of REB events, which these XSELs can find. The total number of REB events during the studied period is 514. Instead of using the closeness of origin times, which might be highly biased for mislocated REB events, as a measure of closeness we checked the difference of arrival times. When three (REB 3) or two (REB 2) from the involved seven stations show the difference less than 4 s we consider this REB event as matched by an XSEL event.

Even after the conflict resolution, the XSELs contain many events close in origin time. We deliberately retain the events in coda of large aftershocks for monitoring purposes. Potential violators of the CTBT may use the coda waves of large earthquakes to hide seismic signals from underground nuclear tests. The XSEL events built in coda may have arrivals less than 10 s far those from the aftershocks. Therefore, some of the REB events were actually matched by two XSEL events. One has three different measures of the number of XSEL events, which found an REB event: FOUND REB – is the total number of XSEL events matching at least one REB event, DOUBLE – is the number of REB events found by two XSEL events, and UNIQUE – the number of matched REB events.

Table 2 demonstrates that the number of matched REB events and arrivals does not differ much between three XSELs despite the number of events in the REAL and SGM XSELs differ by 30%. This might be an indication of a more effective performance of the SGM, which has found almost 98% of the REB events with two and more stations (REB 2) and 80% with three and more stations (REB 3). The REB is not a perfect bulletin and likely misses from 50% to 70% REB ready events, which were successfully found by cross correlation. Therefore, the XSEL events not matched in the REB are not necessary bogus events. At least 400 of them may be valid REB events. Many others are seismically sound but do not match the EDC. For example, have valid arrivals at only two IMS stations.

Among the involved stations, the highest input belongs to MKAR, ZALV, and SONM. Since ASAR and WRA have backazimuths almost opposite to MKAR and ZALV their presence in the XSEL events is crucial for robust local association. It is worth noting that we have limited the number of stations to 7 for the sake of simplicity. For the SGM, it is possible to extend the list of stations to 17 and improve the overall performance of cross correlation pipeline.

Table 3. The number of events in REAL, SGM-REAL, and SGM-GRID XSELs, which match REB events by close (4 s) arrivals at three (REB 3) and two (REB 2) stations; and the number of found REB arrivals at seven stations. The total number of REB events during the studied period is 514.

|  | REB 3 | | | REB 2 | | |
|---|---|---|---|---|---|---|
|  | REAL | SGM-REAL | SGM-GRID | REAL | SGM-REAL | SGM-GRID |
| XSEL | 2324 | 1659 | 1669 | 2324 | 1659 | 1669 |
| FOUND REB | 452 | 439 | 422 | 634 | 616 | 614 |
| DOUBLE | 26 | 33 | 23 | 114 | 112 | 108 |
| UNIQUE | 426 | 406 | 399 | 508 | 504 | 506 |
| ASAR | 163 | 171 | 158 | 230 | 237 | 230 |
| CMAR | 287 | 272 | 251 | 339 | 310 | 293 |
| GERES | 273 | 247 | 239 | 287 | 258 | 255 |
| MKAR | 415 | 362 | 357 | 478 | 425 | 429 |
| SONM | 361 | 328 | 329 | 411 | 383 | 385 |
| WRA | 250 | 254 | 238 | 349 | 339 | 327 |
| ZALV | 367 | 333 | 325 | 398 | 373 | 368 |

4. **Conclusion**

We have assessed the performance of cross correlation with various master events. Currently, the choice of master event meets two major problems: there are no seismic events with high quality signals at many stations to cover even seismically active areas, and it is not feasible to process the real time data flow with a hundred thousand masters after catastrophic earthquakes.

To resolve both major problems it is suggested to replicate waveform templates from the best master event over areas several hundred kilometers in radius.

In this study, we used two principal outputs of cross correlation processing to compare the performance of real masters and a replicated grand master: detections and events. Various characteristics were calculated and relevant distributions were constructed for detection parameters and the number and quality of event hypotheses. These events were built with the same and different sets of stations and the same and different locations. Both, detections and events, have shown that the grand master distributed over a regular grid has the same performance in terms of cross correlation as the real masters. This opens the possibility to populate the broader areas around seismically active zones with replicated grand masters. For the purpose of seismic monitoring, the areas around historical test sites (e.g. Nevada, Aleutian Islands, Semipalatinsk, Novaya Zemlya, Lop Nor, Punggye, Maralinga, and Moruroa) could be populated with nuclear explosions as grand masters. Moreover, any isolated seismic event measured at teleseismic distances by the IMS may serve as a grand master. We are investigating the possibility to use the waveforms from historical nuclear tests detected by non-IMS seismic arrays and 3-C stations as templates for grand masters.

The distribution over a regular grid also reduces the volume of calculations by two orders of magnitude. Overall, an appropriate choice and replication of the best master allows reducing magnitude threshold of seismic monitoring and improving the accuracy and uncertainty of event location at the IDC to the level of the most accurately located events. When a ground truth event is available, one can expand its influence over hundreds of kilometers.

Currently, we are developing for the best method to cover the earth surface with a regular grid of master events. The centre of the area is the original master event. The characteristic length of the grid depends on the distance to the closest stations. In isolated oceanic areas, where the closest array station is at a teleseismic distance, one can use a larger spacing between nodes. Within continents, one or two IMS arrays have to be at regional distances. This is a well-known effect that the level of cross correlation decays much faster with master/slave distance for regional phases (Pg, Pn, Lg) due to stronger velocity fluctuations in the crust relative to the mantle. Therefore, the grid should be denser. However, the change in propagation path makes the replication of a master waveform template almost worthless beyond 20 to 40 km in tectonically active zones with strong lateral inhomogeneity. Continental platforms and shields are more homogeneous. Here we suggest using synthetic seismograms to reshape the master waveforms according to the path change. This concerns only regional phases, however. For teleseismic stations, no difference is expected relative to oceanic basins and a good cross correlation distance is estimated 50 to 100 km.

Aseismic areas can be also covered with high quality waveforms obtained by synthetic methods – from synthetic seismograms to signals synthesized from a large set of actual waveforms observed at a given station, e.g. sub-space detectors (Harris, 2006, 2008; Harris et al, 2006, 2011). For synthetic seismograms, one can use the actual time delays between individual sensors of an array station obtained from regional or even local sources. It is assumed that these delays are of local origin and do not depend on epicentral distance. The use of synthetic seismograms is discussed in Part II of this paper.

**Acknowledgements**

The authors are grateful to all analysts at the IDC for reviewing XSEL and REB events.